\documentstyle[12pt]{article}
\def\be{\begin{eqnarray}}
\def\ee{\end{eqnarray}}
\def\ba{\begin{array}}
\def\ea{\end{array}}

\begin{document}

\begin{center}
{\LARGE {New matrix formalism for\\
\vskip 0.2cm
heterotic string theory 
\vskip 0.2cm
on a torus
}}
\end{center}

\vskip 1.5cm

\begin{center}
{\bf \large {Oleg V. Kechkin}}
\end{center}

\begin{center}
Institute of Nuclear Physics,\\
M.V. Lomonosov Moscow State University, \\
Vorob'jovy Gory, 119899 Moscow, Russia, \\
e-mail: kechkin@depni.npi.msu.su
\end{center}

\vskip 1.5cm
\begin{abstract}
A new Lagrange formalism based on the use of a single scalar 
$(d+1)\times (d+1+n)$ matrix potential is developed for the low--energy 
heterotic string theory with $n$ \, $U(1)$ gauge fields compactified from 
$d+3$ to $3$ dimensions on a torus. This formalism also includes three pairs 
on--shell defined scalar and vector matrix potentials of the dimensions 
$(d+1)\times (d+1)$, $(d+1)\times (d+1+n)$ and $(d+1+n)\times (d+1+n)$. All 
these potentials undergo linear transformations when the group of charging 
symmetries acts.  
\end{abstract}

\newpage

\section{Introduction}

A successful choice of the variables often plays the crucial role in the 
study of a nonlinear theory. The most powerful tools for this study are
related to use of the symmetry methods. A form of the finite symmetry 
transformations depends on the variables choice; the simplest and the 
most convenient symmetry representation has the linear form. This form
allows one to deal with the general group transformation and to obtain the
results which cannot be generalized again by the help of additional
transformations from this symmetry group. In contrast to this situation,
often in real practice one needs in the series of special transformations
to obtain more or less general result in the case of the nonlinear symmetry 
representation. Moreover, at the end of this step--by--step
special symmetry applications it is necessary to establish what kind of 
generality had been obtained.    

In this paper we continue to study the heterotic string theory using the 
symmetry based methods. At low energies this theory becomes the nonlinear
field theory of its massless modes living in the multidimensional
space--time \cite{k}. This effective field theory becomes the nonlinear 
$\sigma$--model after its compactification to three dimensions on a torus
\cite{ms}--\cite{s}. 
Below we develop a new representation for this $\sigma$--model based on 
the use of a single scalar matrix potential which has the lowest
possible matrix dimension. This
scalar matrix potential transforms linearly under the action of the group
of charging symmetries, which preserve trivial spatial asymptotics of the
all $\sigma$--model fields \cite{cs}. In fact, this representation
provides 
the 
compact and convenient tool for the study of asymptotically flat fields and 
for the generation of asymptotically flat solutions of the heterotic string 
theory.      

The paper is organized as follows: in the next section we ``derive'' 
one special $\sigma$--model which can be considered as the simplest
nonlinear  modification of the linear theory, possessing the same charging
symmetry group of transformations. Both the linear and the nonlinear
$\sigma$--models are given in terms of the single scalar nonquadratic matrix 
potential. In the next section we show how the same nonlinear 
$\sigma$--model arises in the framework of heterotic string theory. After 
that, in the next section, we establish the relation between our and chiral 
matrix formalisms. In fact we express the chiral matrix in terms of the 
``main'' scalar matrix potential found in the previous section. Finally, 
in the last section, starting from the conserving chiral current we
construct on shell three pairs of the scalar and vector matrix potentials 
defined in terms of the ``main'' scalar matrix potential. It is shown that 
all these scalar and vector potentials transform linearly under the action 
of charging symmetry group of transformations. In ``Conclusions'' we discuss 
the results obtained and their possible applications to the problem of 
generation of new exact solutions of the heterotic string theory.     


\section{$\sigma$--model derivation}

Let us consider a class of $\sigma$--models possessing two commuting groups 
of symmetry transformations isomorphic to $O(p_1,q_1)$ and $O(p_2,q_2)$
as the ``underlying'' symmetry. Let us choose the real matrix potential 
${\cal Z}$ of the dimension $(p_1+q_1)\times (p_2+q_2)$ as the dynamical 
variable for these $\sigma$--models. Then, the underlying symmetry can be 
realized as the linear transformation
\be \label{1}
{\cal Z} \rightarrow {\cal C}_1{\cal Z}{\cal C}_2, 
\ee
where 
\be \label{2}
{\cal C}_a^T\Sigma_a{\cal C}_a=\Sigma_a, \quad a=1,2, 
\ee
and $\Sigma_a$ are the diagonal matrices with $p_a$ ``$+1$'' and $q_a$ 
``$-1$'' on their main diagonals. Thus, ${\cal C}_a\in O(p_a,q_a)$, where 
$a=1,2$.

Our class of $\sigma$--models is not empty; the Lagrangian of its simplest    
representative has the form
\be \label{3}
L_0=K\,Tr\left ( \Sigma_1\nabla {\cal Z}\,\,\Sigma_2\nabla {\cal Z}^T\right ),
\ee
where $K=const$. This simplest theory is linear, its Euler--Lagrange equation 
is the Laplace equation for the matrix potential ${\cal Z}$:
\be \label{4}
\nabla^2 {\cal Z}=0.
\ee
A set of the nonlinear modifications of the theory (\ref{3}) can be obtained 
using the replacement $\Sigma_a \rightarrow\hat\Sigma_a^{-1}$, 
where the matrix fields $\hat\Sigma_a({\cal Z})$ must transform accordingly 
to the relations
\be \label{5}
\hat\Sigma_1\rightarrow {\cal C}_1\hat\Sigma_1\hat{\cal C}_1^T\qquad
{\rm and}\qquad
\hat\Sigma_2\rightarrow {\cal C}_2^T\hat\Sigma_2\hat{\cal C}_2
\ee 
when the transformation (\ref{1})--(\ref{2}) acts. In this case the 
Lagrangian
\be \label{6}
L=K\,Tr\left ( \hat\Sigma_1^{-1}\nabla {\cal Z}\,\,\hat\Sigma_2^{-1}
\nabla {\cal Z}^T\right )
\ee
remains invariant under the action of the transformation (\ref{1})--(\ref{2}), 
i.e., it belongs to the class under consideration.

It is not difficult to establish the explicit realization of the matrix
fields $\hat\Sigma_a({\cal Z})$. The result can be expressed in terms of
the following formal series:
\be \label{7}
\hat\Sigma_1&=&\kappa_{10}\Sigma_1+\kappa_{11}{\cal Z}\Sigma_2{\cal Z}^T+
\kappa_{12}{\cal Z}\Sigma_2{\cal Z}^T\cdot\Sigma_1\cdot
{\cal Z}\Sigma_2{\cal Z}^T+\cdots,
\nonumber
\\
\hat\Sigma_2&=&\kappa_{20}\Sigma_2+\kappa_{21}{\cal Z}^T\Sigma_1{\cal Z}+
\kappa_{22}{\cal Z}^T\Sigma_1{\cal Z}\cdot\Sigma_2\cdot
{\cal Z}^T\Sigma_1{\cal Z}+\cdots.
\ee 
Here the coefficients $\kappa_{ak}=\kappa_{ak}({\cal Z})$\, $(k=0,1,...)$ 
are the invariants of the transformation (\ref{1})--(\ref{2}); they can be 
taken as the arbitrary functions of the invariants 
$I_l=Tr[(\Sigma_1{\cal Z}\Sigma_2{\cal Z}^T)^l]$, $l=1,2,...$. Thus, we 
have obtained the infinite set of the nonlinear $\sigma$--models possessing 
the ``underlying'' symmetry.

Now our main goal is to choose the single special nonlinear $\sigma$--model 
from this set using some additional principles. In fact we must specify the 
coefficient functions $\kappa_{ak}(\cal Z)$. Our general principle is the 
maximal possible similarity between $L$ and $L_0$, i.e., between the 
nonlinear and the linear $\sigma$--models. We suppose, in particular, that 
$L_0$ is the zero term in the expansion of $L$ into the perturbation series 
with the center at ${\cal Z}=0$. From this it follows that 
$\kappa_{a0}(0)=1$. To obtain more information let us consider the special 
case of $p_1+q_1=p_2+q_2$, when all the matrices under consideration are 
quadratic. We demand $L$ invariance under the inversion
\be \label{7'}
{\cal Z}\rightarrow {\cal Z}^{-1}. 
\ee
This immediately gives $\kappa_{ak}=const$ for any $k$ and, hence, 
$\kappa_{a0}=1$. Moreover, one obtains, that $\kappa_{ak}=0$ for $k\geq 2$, 
and that $\kappa_{11}=\kappa_{21}=\sigma_1 =\pm 1$. Also one obtains that 
$\Sigma_2=\sigma_2\Sigma_1$, where $\sigma_2=\pm 1$. The resulting form of 
the nonlinear Lagrangian $L$ reads:
\be \label{8}
L=K\, Tr\left [ \left ( \Sigma+\sigma{\cal Z}\Sigma{\cal Z}^T\right )^{-1}
\nabla {\cal Z}\left ( \Sigma+\sigma{\cal Z}^T\Sigma{\cal Z}\right )^{-1}
\nabla {\cal Z}^T\right ],
\ee
where $\Sigma =\Sigma_1$, $\sigma=\sigma_1\sigma_2=\pm 1$ and $\sigma_1K$ is 
redefined as $K$. The corresponding form of the linear Lagrangian is
\be \label{9}
L_0=K\, Tr \left [ \Sigma\nabla {\cal Z}\Sigma\nabla {\cal Z}^T\right ].
\ee
It is easy to see that $L_0$ does not possess the inversion (\ref{7'}) as
the symmetry for the arbitrary nondegenerated matrix ${\cal Z}$. However,  
if one considers ${\cal Z}$ restricted by the relation
\be \label{10}
\Sigma{\cal Z}^T\Sigma=c^2{\cal Z}^{-1},
\ee
where $c$ is the arbitrary constant, this symmetry actually takes place. 
This relation means that $c^{-1}{\cal Z}\in O(p_1,q_1)$; the Lagrangians 
$L$ and $L_0$ calculated on this group subspace read: 
\be \label{11}
L=\frac{1}{(1+\sigma c^2)^2}L_0=\frac{c^2K}{(1+\sigma c^2)^2}Tr\left [
\nabla {\cal Z}\nabla {\cal Z}^{-1}\right ].
\ee
Let us demand the numerical identity $L=L_0$ on the group subspace 
(\ref{10}); it is easy to see that it is possible only in the case of 
$c^2=2$ and $\sigma=-1$. This last opportunity fixes the sign in 
Eq. (\ref{8}) as negative. 

The last step is to generalize Eq. (\ref{8}) with $\sigma=-1$ to the 
case of nonquadratic matrix potential ${\cal Z}$. We make it in the 
following natural way:
\be \label{12}  
L=K\, Tr\left [ \left ( \Sigma -{\cal Z}\Xi{\cal Z}^T\right )^{-1}
\nabla {\cal Z}\left ( \Xi-{\cal Z}^T\Sigma{\cal Z}\right )^{-1}
\nabla {\cal Z}^T\right ],
\ee
where
\be \label{13}
\Xi=\left (\begin{array}{crc}
\Sigma& 0\\
0&\tilde\Sigma\\
\end{array}\right )
\ee
and $\tilde\Sigma$ is a diagonal matrix with $\pm 1$ on the main 
diagonal. In this generalization we have supposed, for definiteness,
that $p_2\geq p_1$ and $q_2\geq q_1$; it is easy to see that in the 
special case of $p_2=p_1$ and $q_2=q_1$ we regress to the special 
situation considered above. 

The constructed $\sigma$--model is actually nonlinear: its Euler--Lagrange 
equation can be written in one of two equivalent forms:
\be \label{13'}
\nabla^2{\cal Z}+2\nabla {\cal Z}\Xi {\cal Z}^T
\left ( \Sigma-{\cal Z}\Xi {\cal Z}^T\right )^{-1}\nabla {\cal Z}=0
\ee  
or
\be \label{14}
\nabla^2{\cal Z}+2\nabla {\cal Z}
\left ( \Xi-{\cal Z}^T\Sigma {\cal Z}\right )^{-1}
{\cal Z}^T\Sigma\nabla {\cal Z}=0.
\ee
The matrix field ${\cal Z}$ is defined under the coordinate space $x^{\mu}$. 
Our consideration cannot fix the dimensionality and signature of this 
space--time as well as the value of the parameters $K$, $p_a$ and $q_a$.
All it will be done in the next section where the $\sigma$--model (\ref{12})
will be used for the representation of the toroidally compactified heterotic 
string theory. 


\section{$\sigma$--model in heterotic string theory}

At low energies the bosonic sector of heterotic string theory becomes the
field theory of its massless modes. These modes live in $3+d$--dimensional 
space--time (with the coordinates $X^M,\,\,M=1,...,3+d$) of the signature 
$-+\cdots +$ and include the dilaton field  $\Phi$, the Kalb--Ramond field 
$B_{MN}=-B_{NM}$, the set of $n$ \, $U(1)$ gauge fields $A^I_M$ $(I=1,...,n)$ 
and the metric field (graviton) $G_{MN}=G_{NM}$. The corresponding action  
reads \cite{k}:
\be\label{15}
S_{3+d}=\int d^{3+d}X\sqrt{-det\,G_{MN}}e^{-\Phi} 
\left ( R_{3+d}+\Phi_{,M}\Phi^{,M}-\frac{1}{12}H_{MNK}H^{MNK}-
\frac{1}{4}F^I_{MN}F^{I\,MN}\right ),
\ee
where $H_{MNK}=\partial_MB_{NK}-\frac{1}{2}A^I_MF^I_{NK}+{\rm c.p.\,\, of}
\,\,\{M\,N\,K\}$ and $F^I_{MN}=\partial_MA^I_N-\partial_NA^I_M$. 

Let us perform the toroidal compactification of the first $d$ space--time 
dimensions, i.e., let us consider the fields independent on $X^M$ with 
$M=m=1,...,d$ and possessing the functional dependence on the coordinates 
$X^M$ with $M=d+\mu$, $\mu=1,2,3$. In this case the field components can be 
classified in respect to transformations of the three--dimensional 
coordinates $x^{\mu}\equiv X^{d+\mu}$. Namely, one has
(\cite{ms}--\cite{s}):

\vskip 3mm

\noindent
$1)$ \, three scalar matrices $G$, $B$ and $A$ of the dimensions $d\times d$, 
\,$d\times d$ and $d\times n$ with the components $G_{mk}, B_{mk}$ and 
$A_{mI}=A^I_{m}$ correspondingly, and also the scalar function 
$\phi=\Phi-{\rm ln }\sqrt{-{\rm det}\,G}$; 

\vskip 3mm

\noindent
$2)$ \, three vector columns $\vec A_{1}$, $\vec A_{2}$ and $\vec A_{3}$ of 
the dimensions $d\times 1$, $d\times 1$ and $n\times 1$; their components 
read:
\be \label{17}
(\vec A_1)_{m\mu}&=&(G^{-1})_{mk}G_{k,d+\mu},
\nonumber\\
(\vec A_2)_{m\mu}&=&B_{m,d+\mu}-B_{pq}(\vec A_1)_{m\mu}+
\frac{1}{2}A_{mI}(A_3)_{I\mu},
\nonumber\\
(\vec A_3)_{I\mu}&=&-A^I_{d+\mu}+A^I_m(\vec A_1)_{m\mu} 
\ee
and

\vskip 3mm

\noindent
$3)$ \, two tensor fields
\be \label{18}
h_{\mu\nu}&=&e^{-2\phi}\left [ G_{\mu\nu}-G_{mk}
(\vec A_1)_{m\mu}(\vec A_1)_{k\nu}
\right ],
\nonumber\\
b_{\mu\nu}&=&B_{\mu\nu}-B_{mk}(\vec A_1)_{m\mu}(\vec A_1)_{k\nu}-
\frac{1}{2}\left [(\vec A_1)_{m\mu}(\vec A_2)_{m\nu}-(\vec
A_1)_{m\nu}(\vec A_2)_{m\mu}\right ].
\ee
In three dimensions the field $b_{\mu\nu}$ is nondynamical; following 
\cite{s} we put $b_{\mu\nu}=0$ without contradiction with the motion 
equations. Moreover, in three dimensions it is possible to introduce 
pseudoscalar fields $u$, $v$ and $s$ accordingly the relations 
(\cite{s}, \cite{cs}, \cite{mep})
\be \label{19}
\nabla\times\vec A_1&=&e^{2\phi}G^{-1}
\left [\nabla u+(B+\frac{1}{2}AA^T)\nabla v+A\nabla s\right ],
\nonumber                            \\
\nabla\times\vec A_2&=&e^{2\phi}G\nabla v-
(B+\frac{1}{2}AA^T)\nabla\times\vec A_1+
A\nabla\times\vec A_3.
\nonumber \\
\nabla\times\vec A_3&=&e^{2\phi}
(\nabla s+A^T\nabla v)+A^T\nabla\times\vec A_1,
\ee
Finally, the bosonic sector of the heterotic string theory toroidally
compactified to three dimensions is equivalent on shell to the effective 
three--dimensional theory which describes the interacting scalar fields $G$, 
$B$, $A$ and $\phi$ and the pseudoscalar ones $u$, $v$ and $s$ coupled to 
the metric $h_{\mu\nu}$. 

For our purposes it is useful to embed all the scalar and pseudoscalar 
fields into the following matrices $\cal X$ and $\cal A$ of the 
dimensions $(d+1) \times (d+1)$ and $(d+1) \times n$
correspondingly:
\be \label{20} 
{\cal X}=
\left(
\ba{cc}
-e^{-2\phi}+v^TXv+v^TAs+\frac{1}{2}s^Ts&v^TX-u^T \cr
Xv+u+As&X
\ea
\right),  \qquad
{\cal A}=\left(
\ba{c}
s^T+v^TA \cr
A
\ea
\right),
\ee
where $X=G+B+\frac{1}{2}AA^T$ (\cite{cs}, \cite{mep}). In terms of these 
matrices
the action of the effective three--dimensional theory of heterotic string
reads:
\be \label{21}
S_3=\int d^3x \sqrt h\left \{ -R_3+L_3\right \},
\ee
where
\be \label{22} 
L_3={\rm Tr}\,\left [
\frac{1}{4}\left(\nabla {\cal X}\!-\!\nabla {\cal A}{\cal A}^T
\right)\!{\cal G}^{-1}
\!\left(\nabla {\cal X}^T\!-\!{\cal A}\nabla {\cal A}^T\right)
\!{\cal G}^{-1}
\!+\!\frac{1}{2}\nabla {\cal A}^T{\cal G}^{-1}\nabla {\cal A}
\right ],
\ee
\be \label{23}
{\cal G}=\frac{1}{2}\left({\cal X}+{\cal X}^T-{\cal A}{\cal A}^T\right),
\ee
$h={\rm det}\,h_{\mu\nu}$ and the scalar curvature $R_3$ is constructed using 
the metric $h_{\mu\nu}$. This action describes the nonlinear $\sigma$--model 
coupled to the three--dimensional gravity and has the form of the action
of the stationary Einstein--Maxwell theory \cite{ex}. In
Eq. (\ref{22}) the matrices ${\cal X}$ and ${\cal A}$ play the same role
as the complex Ernst potentials in Einstein--Maxwell theory \cite{e} and
can be named as ``matrix Ernst potentials'' \cite{mep}. 

The main statement of this chapter is the following: the heterotic string
theory $\sigma$--model given by the Lagrangian $L_3$ is the special one from 
the class of $\sigma$--models given by Eq. (\ref{12}). Actually, let us put 
in Eq. (\ref{12}) $K=1$, $\Sigma={\rm diag} \, (-1,-1,+1,\cdots,+1)$ (two 
``$-1$'' and $d-1$ ``$+1$'' on the main diagonal, so $\Sigma$ is the 
$(d+1)\times (d+1)$ matrix) and $\tilde\Sigma=1_n$ (where $1_n$ is the 
$n\times n$ unit matrix). Let us also introduce for the 
$(d+1)\times (d+1+n)$ matrix ${\cal Z}$ the following parametrization:   
\be \label{24}
{\cal Z}=({\cal Z}_1,\,\,\,{\cal Z}_2), 
\ee
where the $(d+1)\times (d+1)$ and $(d+1)\times n$ matrices ${\cal Z}_1$
and ${\cal Z}_2$ read:
\be \label{25}
{\cal Z}_1=2\left({\cal X}+\Sigma\right)^{-1}-\Sigma, 
\qquad {\cal Z}_2=\sqrt 2\left({\cal X}+\Sigma \right)^{-1}{\cal A}
\ee
(the Einstein--Maxwell analogy of these potentials one can find in
\cite{m}).
Then, after some nontrivial algebraical calculations one can prove that
in this special case $L=L_3$. Thus, 
\be \label{26}  
L_3={\rm Tr}\left [ \left ( \Sigma -{\cal Z}\Xi{\cal Z}^T\right )^{-1}
\nabla {\cal Z}\left ( \Xi-{\cal Z}^T\Sigma{\cal Z}\right )^{-1}
\nabla {\cal Z}^T\right ],
\ee 
where the matrices $\Sigma$ and $\Xi$ (see Eq. (\ref{13})) are defined as 
it was explained above. We see that the heterotic string theory 
$\sigma$--model can be represented in terms of the matrix potential 
${\cal Z}$. As the straightforward consequence of this fact we conclude
(see 
Eqs. (\ref{1})--(\ref{2})) that $L_3$ is invariant under the transformation
\be \label{27}
{\cal Z}\rightarrow {\cal Z}^{'}=C_1{\cal Z}C_2,
\ee
where the matrices $C_1$ and $C_2$ satisfy the $O(2,d-1)$ and $O(2,d-1+n)$ 
group relations
\be \label{28}
C_1^T\Sigma C_1=\Sigma \quad {\rm and} \quad C_2^T\Xi C_2=\Xi. 
\ee
Then, the matter part of the heterotic string theory equations is given by 
Eq. (\ref{13'}) or (\ref{14}), whereas the Einstein equation reads:
\be \label{29}
R_{3\,\,\mu\nu}=
{\rm Tr}\left [ \left ( \Sigma -{\cal Z}\Xi{\cal Z}^T\right )^{-1}
\nabla_{(\mu} {\cal Z}\left ( \Xi-{\cal Z}^T\Sigma{\cal Z}\right )^{-1}
\nabla_{\nu)} {\cal Z}^T\right ].
\ee

In \cite{cs} it was shown that Eqs. (\ref{27})--(\ref{28}) gives the total 
group of charging symmetries for the theory under consideration. The
charging symmetries form the subgroup of the complete symmetry group of the 
$\sigma$--model; the transformations from this subgroup preserve trivial 
values of the all three--dimensional ``matter'' fields. These trivial values 
are defined as the zero ones for the fields $B$, $A$, $\phi$, $u$, $v$, $s$ 
and as the $d\times d$ matrix ${\rm diag }(-1,+1,\cdots,+1)$ for $G$. 
From the $(d+3)$--dimensional point of view this field configuration 
(together with $h_{\mu\nu}=\delta_{\mu\nu}$) describes the empty Minkowskian 
space--time in the Carthesian coordinates. From Eq. (\ref{20}) it follows
that
${\cal X}=\Sigma$, ${\cal A}=0$ for the trivial fields and, therefore
(see Eqs.
(\ref{24})--(\ref{25})), ${\cal Z}=0$ in this case. The zero 
${\cal Z}$--value is preserved by the transformation (\ref{27}) with the 
arbitrary matrices $C_1$ and $C_2$; however only the transformation with 
$C_1$ and $C_2$ restricted by the relations (\ref{28}) belongs to the group 
of $\sigma$--model symmetries.

A class of the asymptotically flat solutions of the theory under
consideration is
defined as the class containing all solutions trivial at the spatial 
infinity \cite{cs}, \cite{ex}, \cite{csemda}. Thus, at the spatial
infinity all the solutions from this class
become the trivial field configuration discussed above. It is clear that the 
charging symmetry subgroup contains all the transformations which can be 
applied for generation of the new asymptotically flat solutions from the known 
ones. The noncharging symmetries change the three--dimensional field 
asymptotics and provide trivial action on the solutions; so the charging 
symmetry subgroup includes all the resultative symmetry transformations.  
The linear form of the transformation (\ref{27}) extremely simplifies the 
solution generation procedure; its straightforward application allows one to 
obtain the most general possible generation results for the asymptotically 
flat solutions. To compare our solution generation approach with the
previously known ones one can use \cite{y}.  


\section{Chiral matrix and related structures}

As it has been established above, the effective three--dimensional 
$\sigma$--model of the heterotic string theory compactified on a torus can be 
represented in two different but algebraically related forms: in terms of
the 
potential matrix pair $\cal {X},\,{\cal A}$ and in terms of the single matrix 
potential ${\cal Z}$. Now we consider the chiral matrix representation of the 
same theory, which can be originally given in terms of the pair 
$\cal {X},\,{\cal A}$. Actually, it is easy to prove that the 
$[2(d+1)+n]\times [2(d+1)+n]$ matrix
\be \label{30}
{\cal N}=
\left(
\ba{ccc}
{\cal G}^{-1}&{\cal G}^{-1}{\cal X}&{\cal G}^{-1}{\cal A}\cr
{\cal X}^T{\cal G}^{-1}&{\cal X}^T{\cal G}^{-1}{\cal X}&{\cal X}^T{\cal G}^{-1}{\cal
A}\cr
{\cal A}^T{\cal G}^{-1}&{\cal A}^T{\cal G}^{-1}{\cal X}&{\cal A}^T{\cal G}^{-1}{\cal A}
\ea
\right), 
\ee
satisfies the relation
\be \label{31}
{\cal N}{\cal L}{\cal N}=2{\cal N},
\ee
where the matrix ${\cal L}$ reads
\be \label{32}
{\cal L}=
\left (
\ba{ccc}
0&1&0\cr
1&0&0\cr
0&0&-1
\ea
\right).
\ee
Then the matrix 
\be \label{33}
{\cal M}={\cal N}-{\cal L}
\ee  
satisfies the relations
\be \label{34}
{\cal M}^T{\cal L}{\cal M}={\cal L}, \qquad {\cal M}^T={\cal M}
\ee
which mean that ${\cal M}\in O(d+1,d+1+n)/O(d+1)\times O(d+1+n)$. 
The complet list of $\sigma$--models possessing the matrix 
representations
one can find in \cite{b}. The coset matrix representation for the
heterotic string theory compactified to three dimensions on a torus was   
established at the first time by A. Sen in \cite{s}. Our matrix ${\cal M}$
differs from the Sen's one and related to the matrix Ernst potential 
formulation.

The 
Lagrangian $L_3$ can be rewritten in terms of this coset matrix in the
following ``chiral form'' (\cite{s}, \cite{mep}):
\be \label{35}
L_3=\frac{1}{8}{\rm Tr}(\vec J^2),
\ee
where 
\be \label{36}
\vec J=\nabla {\cal M}\,{\cal M}^{-1}.
\ee
From this it immediately follows that the total symmetry group for
the theory is $O(d+1, d+1+n)$, because $L_3$ remains invariant under the 
action of the transformation
\be \label{37} 
{\cal M}\rightarrow {\cal M}^{'}=\hat{\cal C}^T {\cal M}\hat{\cal C},
\ee
where 
\be \label{38}
\hat{\cal C}^T{\cal L}\hat{\cal C}={\cal L},
\ee
i.e. if $\hat{\cal C}\in O(d+1,d+1+n)$. The group of charging symmetries
forms a subgroup of this total symmetry group; the symmetry matrix 
$\hat{\cal C}$ belongs to this subgroup if it preserves the trivial field
configuration defined in the previous section. This trivial field 
configuration corresponds to the matrix 
\be \label{39}
{\cal M}_0=
\left(
\ba{ccc} 
\Sigma&0&0\cr
0&\Sigma&0\cr   
0&0&1
\ea   
\right),
\ee
so the charging symmetry matrix $\hat{\cal C}$ must satisfy the restriction
\be \label{40}
\hat{\cal C}^T{\cal M}_0 \hat{\cal C}={\cal M}_0
\ee
in addition to the relation (\ref {38}). Eq. (\ref{40}) shows that this
matrix also belongs to the group $O(2, 2d+n)$, so the matrix of
the charging symmetry subgroup
belongs to the intersection of the groups $O(d+1,d+1+n)$ and
$O(2, 2d+n)$. We have the explicit realization
of this symmetry in terms of the potential ${\cal Z}$ (see Eqs. 
(\ref{27})--(\ref{28})). To obtain its chiral realization  
(in the form of Eq. (\ref{37}) with $\hat{\cal C}$ restricted by Eqs.
(\ref{38}) and (\ref{40}))
it is useful to express the chiral matrix ${\cal M}$ in terms of
the matrix potential ${\cal Z}$. To do it, let us introduce the 
$(d+1)\times (d+1+n)$ matrix 
\be \label{41}
{\cal Y}=({\cal X},\,{\cal A}).
\ee  
Then, for the matrix ${\cal N}$ one has:
\be \label{42}             
{\cal N}=
\left(
\ba{cc}
{\cal G}^{-1}&{\cal G}^{-1}{\cal Y}\cr 
{\cal Y}^T{\cal G}^{-1}&{\cal Y}^T{\cal G}^{-1}{\cal Y}
\ea
\right).
\ee     
Here the matrix ${\cal G}$ can also be written by the help of ${\cal Y}$:
\be \label{43} 
{\cal G}=\frac{1}{2}\left [
{\cal Y}l^T+l{\cal Y}+{\cal Y}(1-l^Tl){\cal Y}^T
\right ],
\ee
where the matrix
\be \label{44}
l=(1\,\,0)
\ee
is constructed from the $(d+1)\times (d+1)$ and $(d+1)\times n$ blocks. 
Now one can express the matrices ${\cal Y}$ and ${\cal G}$ in terms 
of the matrix potential ${\cal Z}$ using Eqs. (\ref{24}) and (\ref{25});
the result is:
\be \label{45}
{\cal Y}=(\Sigma+{\cal Z}l^T)^{-1}(l+{\cal Z}P), \quad
{\cal G}=(\Sigma+{\cal Z}l^T)^{-1}(\Sigma-{\cal Z}\Xi {\cal Z}^T)
(\Sigma+l{\cal Z}^T)^{-1},
\ee  
where 
\be \label{46}
P=
\left(
\ba{cc}   
-\Sigma&0\cr
0&\sqrt 2
\ea
\right).
\ee
Using these relations, one establishes that 
\be \label{47}  
{\cal N}=\Psi^TH^{-1}\Psi,
\ee
where $H=\hat\Sigma_1
=\Sigma-{\cal Z}\Xi {\cal Z}^T
$ and $\Psi=\Psi ({\cal Z})$ is the linear $(d+1)\times [2(d+1)+n]$ matrix 
function
\be \label{49}
\Psi={\cal A}+{\cal Z}{\cal B}
\ee
with the proportionality coefficient matrices
\be \label{50}
{\cal A}=(\Sigma,\,l),\qquad {\cal B}=(l^T,\,P).
\ee
Thus, for the explicit form of ${\cal M}({\cal Z})$ one obtains:
\be \label{51}
{\cal M}=({\cal A}+{\cal Z}{\cal B})^TH^{-1}({\cal A}+{\cal Z}{\cal B})
-{\cal L}.
\ee

As it can be easily verified, the matrices ${\cal A}$ and ${\cal B}$:
yield the following remarkable properties:
\be \label{52}
&(a)& \quad {\cal A}{\cal L}{\cal A}^T=2\Sigma, \nonumber\\
&(b)& \quad {\cal A}{\cal L}{\cal B}^T=0, \nonumber\\
&(c)& \quad {\cal B}{\cal L}{\cal B}^T=-2\Xi.
\ee
Now using them and Eq. (\ref{51}) we are ready to calculate the matrices
$\hat{\cal C}_a$ ($a=1,2$) which define the transformations (\ref{37}) of 
the chiral matrix ${\cal M}$ corresponding to the ones of the potential 
${\cal Z}$ given by the matrices ${\cal C}_a$ (see Eqs. (\ref{27}) and 
(\ref{28})). Actually, from Eqs. (\ref{38}) and (\ref{51}) it follows
that
\be \label{53}
{\cal M}^{'}=\hat{\cal C}_a^T\left [
({\cal A}+{\cal Z}{\cal B})^TH^{-1}({\cal A}+{\cal Z}{\cal B})
-{\cal L}
\right ]\hat{\cal C}_a;
\ee
whereas from Eqs. (\ref{27}) and (\ref{51}) one obtains that
\be \label{54}
{\cal M}^{'}=
({\cal A}+{\cal Z}^{'}{\cal B})^TH^{'\,\,-1}({\cal A}+{\cal Z}^{'}{\cal B})
-{\cal L}, 
\ee
where
\be \label{55}
{\cal Z}^{'}={\cal C}_1{\cal Z} \quad {\rm and} \quad
{\cal Z}^{'}={\cal Z}{\cal C}_2
\ee
for $a=1$ and $a=2$ correspondingly. From comparison of Eq. (\ref{53}) and 
Eqs. (\ref{54})--(\ref{55}) it is easy to obtain the relations defining 
the matrices $\hat{\cal C}_a$. So, for $a=1$ one has 
($H^{'}={\cal C}_1H{\cal C}_1^T$)
\be \label{56}
&(a)& \quad {\cal B}\hat{\cal C}_1={\cal B},\nonumber\\
&(b)& \quad {\cal A}\hat{\cal C}_1={\cal C}_1^{-1}{\cal A},\nonumber\\
&(c)& \quad \hat{\cal C}_1^T{\cal L}\hat{\cal C}_1={\cal L}. 
\ee
From Eqs. (\ref{56}-a) and (\ref{52}-b) it follows that 
$\hat{\cal C}_1=1+{\cal L}{\cal A}^T\xi$, where the matrix 
$\xi=\frac{1}{2}\Sigma ({\cal C}_1^{-1}-1){\cal A}$ in view of Eq. 
(\ref{56}-b). Thus, the matrix $\hat{\cal C}_1$ reads:
\be \label{57}
\hat{\cal C}_1
=1+\frac{1}{2}{\cal L}{\cal A}^T\Sigma ({\cal C}_1^{-1}-1){\cal A};
\ee  
the condition (\ref{56}-c) is satisfied in view of Eqs. (\ref{28}) and 
(\ref{52}-a). Then, for $a=2$ the relations defining $\hat{\cal C}_2$ 
relations read ($H^{'}=H$):
\be \label{58}
&(a)& \quad {\cal A}\hat{\cal C}_2={\cal A},\nonumber\\
&(b)& \quad {\cal B}\hat{\cal C}_2={\cal C}_2{\cal B},\nonumber\\
&(c)& \quad \hat{\cal C}_2^T{\cal L}\hat{\cal C}_2={\cal L}.
\ee
From Eqs. (\ref{58}-a) and (\ref{52}-b) it follows that
$\hat{\cal C}_1=1+{\cal L}{\cal B}^T\eta$, where the matrix
$\eta=\frac{1}{2}\Xi (1-{\cal C}_2){\cal B}$ in view of Eq. (\ref{58}-b).
Thus, the matrix $\hat{\cal C}_2$ reads:
\be \label{59}
\hat{\cal C}_2=1+\frac{1}{2}{\cal L}{\cal B}^T\Xi (1-{\cal C}_2){\cal B};
\ee
the condition (\ref{58}-c) is satisfied in view of Eqs. (\ref{28}) and
(\ref{52}-c).
It is easy to see that from Eq. (\ref {52}-b) it also follows that 
\be \label{60}
[\hat{\cal C}_1,\,\hat{\cal C}_2]=0, 
\ee
i.e., the charging symmetry subgroups $`1'$ and $`2'$ commute; this fact also
follows from Eq. (\ref{27}) which gives the action of the same subgroups in 
the ${\cal Z}$--representation. Moreover, from Eqs. {\ref{57}} and (\ref{59}) 
it follows that the matrices ${\cal A}$ and ${\cal B}$ provide the 
transition from the ${\cal Z}$--representation to the chiral one for the 
charging symmetry subgroups $`1'$ and $`2'$ correspondingly. 


\section{Vector matrix potentials}

It is possible to express the Lagrangian $L_3$ only in terms of the matrix 
function $\Psi$. Actually, using the relations (\ref{52}) one can prove
that 
\be \label{61}
H=\frac{1}{2}\Psi{\cal L}\Psi^T,
\ee
so
\be \label{62}
{\cal M}=2{\Psi}^T(\Psi{\cal L}\Psi^T)^{-1}\Psi-{\cal L}.
\ee
Then, using the relations (\ref{34}) and (\ref{36}) it is easy to show that
\be \label{63} 
\vec j=\vec J{\cal L}=\left ( 1-\frac{1}{2}\Psi^TH^{-1} 
\Psi{\cal L}\right )\nabla\Psi^T H^{-1}\Psi-
\Psi^TH^{-1}\nabla\Psi \left ( 1-\frac{1}{2}{\cal L}\Psi^TH^{-1}\Psi\right ),
\ee
and from Eq. (\ref{35}) for the three--dimensional Lagrangian $L_3$ one 
obtains:
\be \label{64}
L_3=-\frac{1}{2}{\rm Tr}\left [\nabla\Psi \left ({\cal L}-
\frac{1}{2}{\cal L}\Psi^TH^{-1}\Psi{\cal L}\right )
\nabla\Psi^T T^{-1}\right ].
\ee
Now, using Eqs. (\ref{49}) and (\ref{52}) it is easy to check that 
Eq. (\ref{26}) actually takes place as the consequence of Eq. (\ref{64})
and, hence, Eqs. (\ref{22}) and (\ref{35}) (it is not difficult to establish 
the equivalence of $L_3$ defined by Eqs. (\ref{22}) and (\ref{35})).

Then, from Eqs. (\ref{37}), (\ref{38}) and (\ref{63}) it follows that the 
transformation low for the matrix current $\vec j$ reads:
\be \label{65}
\vec j\rightarrow \vec j^{'}=\hat{\cal C}^T\vec j\hat{\cal C},
\ee
so it has the same form as the one for the chiral matrix ${\cal M}$. The
matrix current $\vec j
$ is antisymmetric
($\vec j=\nabla{\cal M}\,{\cal L}{\cal M}=-{\cal M}{\cal L}
\nabla{\cal M}=-\vec j^T$);
this property is preserved by the transformation (\ref{65}). In this
section we
study the action of the charging symmetry group of transformations on the 
vector matrix potentials which can be introduced as the potentials related 
to the divergence--free vector field $\vec j$. Namely, from the chiral form 
(\ref{35}) of the Lagrangian $L_3$ it follows that the matter part of motion 
equations reads:
\be \label{66}
\nabla \vec j=0,
\ee
so one can introduce on shell the vector matrix potential $\vec\Omega$
accordingly the relation
\be \label{67}
\nabla\times\vec\Omega=\vec j.
\ee
From Eqs. (\ref{65}) and (\ref{67}) one obtains that the matrix potential
$\vec\Omega$ also transforms under the action of the general symmetry group
as the chiral matrix $\cal M$,
\be \label{68}
\vec\Omega\rightarrow \vec\Omega^{'}=\hat{\cal C}^T\vec\Omega\hat{\cal C}.
\ee
The matrix potential $\vec\Omega$ is antisymmetric, this property is
preserved by the
transformation (\ref{68}). We would like to extract from $\vec\Omega$ the
set of independent vector potentials, to express these potentials in terms 
of ${\cal Z}$ and to establish the action of the charging symmetry group of 
transformations on this set.

To realize this program let us express the current $\vec j$ in terms of 
the potential ${\cal Z}$ using Eqs. (\ref{51})--(\ref{52}); the result
reads:
\be \label{69}
\vec j=-{\cal A}^T\vec j_1{\cal B}+{\cal B}^T\vec j_1^T{\cal A}+
{\cal A}^T\vec j_2{\cal A}+{\cal B}^T\vec j_3{\cal B},
\ee
where
\be \label{70}
\vec j_1&=&H^{-1}\nabla{\cal Z}-H^{-1}\left (
{\cal Z}\,\Xi\nabla{\cal Z}^T-\nabla{\cal Z}\Xi{\cal Z}^T\right )H^{-1}
{\cal Z},\nonumber\\
\vec j_2&=&H^{-1}\left (
{\cal Z}\,\Xi\nabla{\cal Z}^T-\nabla{\cal Z}\Xi{\cal Z}^T\right )H^{-1},
\nonumber\\
\vec j_3&=&\nabla{\cal Z}^TH^{-1}{\cal Z}-{\cal Z}^TH^{-1}\nabla{\cal Z}+
{\cal Z}^TH^{-1}\left (
{\cal Z}\Xi\nabla{\cal Z}^T-\nabla{\cal Z}\Xi{\cal Z}^T\right )H^{-1}
{\cal Z}.
\ee
Thus, the current $\vec j$ is a linear combination of three currents 
$\vec j_i$, \, $i=1,2,3$. These currents are obviously linear independent 
and expressed in terms of ${\cal Z}$; also they are divergent--free. To 
prove this fact, let us introduce the `proection' operators
\be \label{71}
\Pi_1=\frac{1}{2}{\cal L}{\cal A}^T\Sigma \qquad
{\rm and}\qquad
\Pi_2=-\frac{1}{2}{\cal L}{\cal B}^T\Xi. 
\ee 
From the relations (\ref{52}) it follows that 
\be \label{72}
{\cal A}\Pi_1&=&1, \quad {\cal B}\Pi_1=0,\nonumber\\
{\cal A}\Pi_2&=&0, \quad {\cal B}\Pi_2=1,
\ee
so
\be \label{73}
\vec j_1=-\Pi_2^T\vec j\Pi_1,\quad\vec j_2=\Pi_1^T\vec j\Pi_1,\quad
\vec j_3=\Pi_2^T\vec j\Pi_2.
\ee
Thus, from Eq. (\ref{66}) it follows that $\nabla\vec j_i=0$. Let us now 
define three vector matrix potentials $\vec\Omega_i$ accordingly the 
relations
\be \label{74} 
\nabla\times\vec\Omega_i=j_i;
\ee
from Eq. (\ref{70}) one concludes that $\vec\Omega_1$, $\vec\Omega_2$
and $\vec\Omega_3$ have the matrix dimensions $(d+1)\times (d+1+n)$, 
$(d+1)\times (d+1)$ and $(d+1+n)\times (d+1+n)$ correspondingly.

The matrix vector potentials $\vec\Omega_i$ are algebraically independent.
The action of the group of charging symmetry transformations on the set of 
$\vec\Omega_i$ can be established using Eqs. (\ref{57}), (\ref{59}) and 
(\ref{68}). However, the same result can be easily obtained from Eqs. 
(\ref{27}) and (\ref{70}); it reads:
\be \label{75}
\vec\Omega_1\rightarrow\vec\Omega_1^{'}
=
{\cal C}_1^{T\,\,-1}\vec\Omega_1{\cal C}_2,
\quad 
\vec\Omega_2\rightarrow\vec\Omega_2^{'}
= 
{\cal C}_1^{T\,\,-1}\vec\Omega_2{\cal C}_1^{-1},
\quad 
\vec\Omega_3\rightarrow\vec\Omega_3^{'}
= 
{\cal C}_2^T\vec\Omega_3{\cal C}_2.
\ee
Thus, the charging symmetry transformations do not `mix' the potentials
$\vec\Omega_i$, and these vector potentials (as well as the scalar one 
${\cal Z}$) transform as singlets of the group of charging symmetry 
transformations.  

Now let us note that the matrices ${\cal M}$ and $\vec \Omega$ has the very 
similar structure in terms of their linearly independent components. 
Actually, for the matrix $\vec\Omega$ one has (see Eqs. (\ref{67}) and 
(\ref{69})):
\be \label{76}
\vec\Omega=-{\cal A}^T\vec\Omega_1{\cal B}+{\cal B}^T\vec\Omega_1^T{\cal A}+
{\cal A}^T\vec\Omega_2{\cal A}+{\cal B}^T\vec\Omega_3{\cal B},
\ee
whereas for the matrix ${\cal M}$ one obtains from Eq. (\ref{51}) that
\be \label{77}
{\cal M}={\cal A}^T{\cal M}_1{\cal B}+{\cal B}^T{\cal M}_1^T{\cal A}+
{\cal A}^T{\cal M}_2{\cal A}+{\cal B}^T{\cal M}_3{\cal B}-{\cal L},
\ee
where
\be \label{78}
{\cal M}_1
=
H^{-1}{\cal Z},
\quad
{\cal M}_2
=
H^{-1},
\quad 
{\cal M}_3
=
{\cal Z}^TH^{-1}{\cal Z}.
\ee  
Both the sets $\vec\Omega_i$ and ${\cal M}_i$ can be extracted from their 
linear combinations $\vec\Omega$ and ${\cal M}$ (Eqs. (\ref{76}) and 
(\ref{77})) in the same manner as it had been done for the currents
$\vec j_i$
extracted from Eq. (\ref{69}). Thus, for the components ${\cal M}_i$ 
one has:
\be \label{79}
{\cal M}_1=\Pi_1^T{\cal M}\Pi_2,\quad{\cal M}_2=\Pi_1^T{\cal
M}\Pi_1+\frac{1}{2}\Sigma,\quad
{\cal M}_3=\Pi_2^T{\cal M}\Pi_2-\frac{1}{2}\Xi
\ee
(where the relations 
\be \label{80}
\Pi_1^T{\cal L}\Pi_2=0,\qquad \Pi_1^T{\cal L}\Pi_1=\frac{1}{2}\Sigma,\qquad 
\Pi_2^T{\cal L}\Pi_2=-\frac{1}{2}\Xi
\ee
following from Eq. (\ref{52}) had been used).

It is easy to see that the matrices ${\cal M}_i$ transform exactly as the 
matrices $\vec\Omega_i$ under the action of charging symmetry group, i.e., 
\be \label{80'}
{\cal M}_1\rightarrow{\cal M}_1^{'}
=
{\cal C}_1^{T\,\,-1}{\cal M}_1{\cal C}_2,
\quad 
{\cal M}_2\rightarrow{\cal M}_2^{'}
= 
{\cal C}_1^{T\,\,-1}{\cal M}_2{\cal C}_1^{-1},
\quad 
{\cal M}_3\rightarrow{\cal M}_3^{'}
= 
{\cal C}_2^T{\cal M}_3{\cal C}_2.
\ee
Thus, we have established two sets of the linearly independent matrix
potentials,
the scalar ${\cal M}_i$ and the vector $\vec\Omega_i$ ones. These scalar and 
vector potentials with equal indeces have the same matrix dimensions and the 
same transformation properties. They can be combined into three pairs of 
potentials with one scalar and one vector matrix potentials in any pair. All
these potentials are defined in terms of the ``main'' matrix potential
${\cal Z}$; using Eq. (\ref{67}) it is possible to relate vector and scalar
matrix potentials.

At the end of this chapter let us come back to the linear theory described by 
the Lagrangian $L_0$. The heterotic string theory $\sigma$--model had been 
constructed as some ``minimal'' nonlinear generalization of this linear 
theory, so one can wait that the properties of the nonlinear theory, being 
linearized in the appropriate way, take place for the linear one. WE would \
like to establish the sets of the potentials ${\cal M}_i$ and
$\vec\Omega_i$ for the linear theory. To do it let us replace 
\be \label{81}
H=\hat\Sigma_1\rightarrow \Sigma_1=\Sigma. 
\ee
This immediately gives the following relations for
the components ${\cal M}_i$ (see Eq. (\ref{78})):
\be \label{82}
{\cal M}_1
=
\Sigma{\cal Z},
\quad 
{\cal M}_2
=
\Sigma,
\quad 
{\cal M}_3&=&{\cal Z}^T\Sigma{\cal Z}.
\ee
It is easy to see that the charging symmetry transformations act on this
set of ${\cal M}_i$ exactly as on the nonlinear one. Then, if one also 
removes the nonlinear terms of the third and fourth square from 
Eq. (\ref{70}), one obtains the set of matrix currents
\be \label{83}
\vec j_1&=&\Sigma\nabla{\cal Z}
\nonumber\\
\vec j_2&=&\Sigma\left (
{\cal Z}\,\Xi\nabla{\cal Z}^T-\nabla{\cal Z}\Xi{\cal Z}^T\right )\Sigma,
\nonumber\\
\vec j_3&=&\nabla{\cal Z}^T\Sigma{\cal Z}-{\cal Z}^T\Sigma\nabla{\cal Z}
\ee
which conserve on shell (i.e., have the zero divergences), when the motion 
equation (\ref{4}) of the linear theory is satisfied. After that we define 
the set of $\vec\Omega_i$ accordingly Eq. (\ref{74}) and see, that these 
vector potentials also transform as their analogies constructed for the 
heterotic string theory. We can go even more far and to define the matrices 
$\vec j$, ${\cal M}$ and $\vec\Omega$ for the linear theory 
accordingly Eqs. (\ref{69}), (\ref{76}) and (\ref{77}), where the sets 
$\vec j_i$, ${\cal M}_i$ and $\vec\Omega_i$ must be taken in their 
linearized form. It is clear that, for example, Eq. (\ref{66}) takes place
again, but now ${\cal L}{\cal M}{\cal L}\neq{\cal M}^{-1}$ and 
$\nabla{\cal M}\,{\cal M}^{-1}{\cal L}\neq\vec j$. Thus, it is impossible 
to rewrite the linear theory (\ref{3}) in the same chiral form as the
heterotic string theory. However this fact is not really surprising, 
because in the opposite case one can make the false conclusion about 
the equivalence of these essentially different theories. 


\section{Conclusion}

In this paper we have developed a new formalism for the heterotic string
theory compactified to three dimensions on a torus. This formalism is based 
on the use of the single real matrix potential ${\cal Z}$, which has the 
lowest possible matrix dimension possessing by the theory. Actually, 
${\cal Z}$ is the $(d+1)\times (d+1+n)$ matrix, whereas the number of 
components of the scalar fields for the effective three--dimensional 
heterotic string theory is exactly $(d+1)\cdot (d+1+n)$ (see Eq. (\ref{20})). 
The new formalism is naturally related to two sets of three scalar and
three 
vector matrix potentials ${\cal M}_i$ and $\vec\Omega_i$
(see some analogies for the case of General Relativity in
\cite{kin}). Both
these sets, 
as well as the potential ${\cal Z}$, undergo linear transformations under 
the action of the charging symmetries.

The new formalism can be effectively applied for generation of the
asymptotically flat solutions of the heterotic string theory. The 
straightforward application of the relations (\ref{75}) and (\ref{78}) 
allows one to obtain the complete set of components for the vector and 
scalar fields of the generated solution using only the algebraical (not 
differential) calculations. The result of this generation procedure will 
be nongeneralizable by means of the three--dimensional charging symmetries 
in the case when one uses for generation the matrices ${\cal C}_a$ 
restricted only by Eq. (\ref{28}). 

The developed formalism can be used for straightforward construction of the
new solutions. In the forthcoming publications we hope to present some 
charging symmetry complete classes of the asymptotically flat solutions 
constructed using the ${\cal Z}$--representation of the theory. Here we 
would like to note that the potentials ${\cal M}_i$ and $\vec\Omega_i$ 
provide the compact and convenient tool for the study of monopole and dipole 
characteristics of any asymptotically flat field configuration. Moreover, 
in this analysis one can use the linearized variants of the matrix 
potentials, because the higher nonlinear terms correspond to the higher 
multipole moments.


\section*{Acknowledgments}

This work was supported by RFBR grant $N^{0} \,\, 00\,02\,17135$.



\begin{thebibliography}{30}

\bibitem {k}
E. Kiritsis, ``Introduction to superstring theory'',
Lectures presented at the Catholic University of Leuven and University 
of Padova, 1996-1997. Leuven, Belgium: Leuven Univ. Pr. (1998) 315 p, 
(Leuven notes in mathematical and theoretical physics. B9). 

\bibitem {ms}
J. Maharana and J.H. Schwarz,
Nucl. Phys. {\bf B390} (1993) 3.


\bibitem{s}
A. Sen, 
Nucl. Phys. {\bf B434} (1995) 179.

\bibitem{cs}
A. Herrera--Aguilar and O.V. Kechkin,
Phys. Rev. {\bf D59} (1999) 124006.

\bibitem{mep}
A. Herrera Aguilar, O.V. Kechkin,
Int. J. Mod. Phys. {\bf A13} (1998) 393.

\bibitem{ex}
D. Kramer, H. Stephani, M. MacCallum, E. Herlt,
``Exact solutions of the Einstein field equations'',
Deutcher Verlag der Wissenschaften, Berlin, 1980.
\bibitem{e}
F. J. Ernst, Phys. Rev. {\bf 168} 5 (1968) 1415;
Phys. Rev. {\bf 167} 5 (1968) 1175.

\bibitem{m}
P.O. Mazur,
Acta Phys. Pol., {\bf B14} (1983) 219.

\bibitem{csemda}
A. Herrera--Aguilar, O.V. Kechkin,
Mod. Phys. Lett. {\bf A13} (1998) 1907.

\bibitem{y}
D. Yuom,
Phys. Rept. {\bf 316} (1999) 1.

\bibitem{b}
P. Breitenlohner, D. Maison G. W. Gibbons 
Commun. Math. Phys. {\bf 120} (1988) 225;
P. Breitenlohner, Dieter Maison 
Commun. Math. Phys. {\bf 209} (2000) 785. 
 

\bibitem{kin}
W. Kinnersley, J. Math. Phys. {\bf 14} (1973) 651.

\end{thebibliography}
\end{document}